\documentclass[12pt,a4paper]{article}

\usepackage{amsmath}
\usepackage{amsfonts}
\usepackage{amssymb}
\usepackage{amsthm}

\newcommand{\C}{\mathbb C}  
\renewcommand{\H}{\mathbb H}            %

\newcommand{\dd}{{\rm d}} 

\theoremstyle{plain} 

\theoremstyle{definition}

\theoremstyle{remark}

\begin{document}

\title{A Lorentzian version of the non-commutative geometry of the standard model of particle physics}

\author{John W. Barrett
\\ \\
School of Mathematical Sciences\\
University of Nottingham\\
University Park\\
Nottingham NG7 2RD, UK\\
\\
john.barrett@nottingham.ac.uk\\
}

\date{31 August 2006, revised 3 November 2006}

\maketitle

\begin{abstract}   A formulation of the non-commutative geometry for the standard model of particle physics with a Lorentzian signature metric is presented. The elimination of the fermion doubling in the Lorentzian case is achieved by a modification of Connes' internal space geometry so that it has signature 6 (mod 8) 
rather than 0. The fermionic part of the Connes-Chamseddine spectral action can be formulated, and it is shown that it allows an extension with right-handed neutrinos and the correct mass terms for the see-saw mechanism of neutrino mass generation.
\end{abstract}

\section{The Euclidean model}

The hidden geometrical structure of the standard model of particle physics was discovered by Connes using non-commutative geometry\cite{C}. His model suffers from two defects from the physical point of view: firstly that the space-time metric is Euclidean, and secondly that each particle appears four times, not once\cite{LMMS, GIS}. The purpose of this paper is to give the analogous geometrical framework for the standard model with Lorentzian signature which also, at the same time, solves the particle quadrupling problem. This model allows the introduction of neutrino masses using the see-saw mechanism.

Connes' formulation of non-commutative geometry is a real spectral triple \cite{CR}. This consists of a Hilbert space $\cal H$, an algebra $\cal A$ of bounded operators in $\cal H$, and the Dirac operator $D$, a self-adjoint operator in $\cal H$, together with a chirality operator $\gamma$ in $\cal H$ and an antilinear map $J$ on $\cal H$ called the real structure. The chirality satisfies $\gamma=\gamma^*$ and $\gamma^2=1$ and the idea is that its two eigenvalues label the left- and right- handed particles, left-handed particles having eigenvalue 1, right-handed -1. This data satisfies a number of axioms, some of which depend on an integer parameter mod 8, the signature of the geometry, $\sigma$. If $\sigma$ is even, these axioms are
$$J^2=\epsilon$$
$$J\gamma=\epsilon''\gamma J$$
where $\epsilon$ and $\epsilon''=\pm1$ are given by the table
$$\begin{matrix}
\sigma&0&2&4&6\\
\epsilon&1&-1&-1&1\\
\epsilon''&1&-1&1&-1\\
\end{matrix}$$
The parameter $\sigma$ was previously identified by Connes as the dimension of the spectral triple. This makes sense for Euclidean geometries, where, by analogy with a Riemannian metric, signature and dimension are the same thing. However for a Lorentzian metric, signature and dimension are different.

The geometry of the Euclidean standard model is a product of a real spectral triple $({\cal H}_M,{\cal A}_M,D_M,\gamma_M,J_M)$ for the space-time manifold $M$ with a finite-dimensional real spectral triple 
$({\cal H}_F,{\cal A}_F,D_F,\gamma_F,J_F)$ for a non-commutative `internal space'. The Hilbert space ${\cal H}_F$ is constructed with basis the 45 elementary fermions, counting left- and right-handed particles separately, and another 45 basis vectors for the antiparticles of the same fermions. Since the space ${\cal H}_M$ is the Hilbert space of Dirac spinors on $M$, the particle quadrupling phenomenon is apparent. The total space of fermions is 
$${\cal H}={\cal H}_M\otimes{\cal H}_F.$$

A given elementary fermion, for example a left-handed electron $e_L\in{\cal H}_F$, determines a subspace of wavefunctions $\phi=\psi\otimes e_L+\psi'\otimes\overline e_L\in{\cal H}$, where $\overline e_L=J_Fe_L$ is the independent basis vector corresponding to the antiparticle, and $\psi$ and $\psi'$ are arbitrary Dirac spinors. In the standard model, $e_L$ should only have a Weyl spinor, so the degrees of freedom are overcounted four-fold. One would like to reduce the space $\cal H$ by imposing the relations \cite{BK}
$$J\phi=\phi$$
and 
$$\gamma\phi=\phi$$
where $J=J_M\otimes J_F$ and $\gamma=\gamma_M\otimes \gamma_F$ are the real structure and chirality for the total space $\cal H$. However this does not make sense; their consistency would require
$$J^2=1$$
and
$$J\gamma=\gamma J$$
which hold only if the total signature is equal to zero. The total signature for the Euclidean model is unfortunately four.

\section{The Lorentzian model}
To construct the Lorentzian signature model it is necessary to replace the spectral triple for the manifold by the analogous structure for a Lorentz-signature space-time. The guiding idea for this paper is that the structure of the internal space is determined by the requirement that it should be possible to eliminate the quadrupling of the fermions by imposing the additional requirement that the physical fields are simultaneously eigenvectors of $J$ and $\gamma$. The result, explained below, is that the internal space is thus determined to be a real spectral triple of signature six. Since it corresponds to a 0-dimensional manifold (having a finite-dimensional Hilbert space), one can see that it is also in some sense Lorentzian.

In a Lorentzian geometry the usual formula for the inner product of spinors
$$(\psi,\psi')=\int\overline\psi\psi'\;\dd^4x$$
(written here for Minkowski space) is indefinite, and so ${\cal H}_M$ is no longer a Hilbert space. The spinors in ${\cal H}_M$ are taken to be smooth functions of compact support, though one could probably accomodate other boundary conditions as required by quantum field theory. The Dirac operator is self-adjoint, i.e.
$$(\psi,D_M\psi')=(D_M\psi,\psi').$$
The chirality operator $\gamma_M$ is also self-adjoint if its eigenvalues are defined as $+i$ for left-handed particles and $-i$ for right-handed ones. Clearly, $\gamma_M^2=-1$. The real structure $J_M$ is charge conjugation.

These obey the relations
$$D_M\gamma_M+\gamma_MD_M=0$$
$$D_MJ_M=J_MD_M$$
$$J_M\gamma_M=\gamma_MJ_M$$
$$J_M^2=1$$
$$(J_M\psi,J_M\psi')=-\overline{(\psi,\psi')}.$$
In addition, compatibility with the algebra of functions ${\cal A}_M$  is expressed by the relations
$$\gamma_M a=a\gamma_M$$
$$J_M a J_M^{-1}=a^*$$
$$[[D,a],b]=0,$$
for all $a,b\in{\cal A}_M$.

Following the guiding principle explained above, the product of this geometry with a real spectral triple having a finite-dimensional Hilbert space is now computed. The signature of this internal space is for the moment arbitrary. The product geometry is
$${\cal H}={\cal H}_M\otimes{\cal H}_F$$
$${\cal A}={\cal A}_M\otimes {\cal A}_F$$
$$\gamma=\gamma_M\otimes\gamma_F$$
$$J=J_M\otimes J_F$$
$$D=D_M\otimes 1+\gamma_M\otimes D_F.$$

A calculation shows that this product geometry obeys the relations
$$D^*=D$$
$$\gamma^*=\gamma,\quad\quad \gamma^2=-1$$
$$D\gamma+\gamma D=0$$
$$DJ=JD$$
$$J\gamma=\epsilon''\gamma J$$
$$J^2=\epsilon$$
$$(J\Psi,J\Psi')=-\overline{(\Psi,\Psi')},$$
where $\epsilon$, $\epsilon''$ are determined by the signature $\sigma$ of the internal space, together with the usual relations for the action of the algebra
$$\gamma a=a\gamma$$
$$[Ja^*J^{-1},b]=0$$
$$[[D,a],Jb^*J^{-1}]=0$$ for all $a,b\in{\cal A}$. The last of these is called the first-order axiom.

Since the eigenvalues of $\gamma$ are now $\pm i$, consistency in applying the additional relations
$$\gamma \Psi=i\Psi$$
$$J\Psi=\Psi$$
to $\cal H$ to define the physical subspace of fields ${\cal K}\subset{\cal H}$ requires that
$$J^2=1$$
$$J\gamma+\gamma J=0,$$
which is satisfied only when $\sigma=6$. Thus the requirements of the standard model can only be satisfied if the internal space is taken to be a finite-dimensional real spectral triple of signature six. 

In this case, one can understand the physical content of the constraints. For an electron field
$\phi=\psi\otimes e_L+\psi'\otimes\overline e_L\in{\cal K}$, they require that $\psi$ is a left-handed
spinor, and that $\psi'=J_M\psi$, the charge conjugate. Thus the physical degrees of freedom are accurately represented.

\section{Modification of the internal geometry}
Such a triple of signature six can be constructed by modifying the construction of Connes. The Hilbert space ${\cal H}_F$ and algebra ${\cal A}_F=\C\oplus\H\oplus M_3(\C)$ are the same, with the same action, but $\gamma_F$ is defined by
$$\gamma_F f_L=f_L$$
$$\gamma_F \overline f_L= - \overline f_L$$
$$\gamma_F f_R=-f_R$$
$$\gamma_F \overline f_R=f_R,$$
using the notation $f_L$ for a left-handed fermion and $\overline f_L$ for the antiparticle. The operator $J_F$ is defined to be
$$ J_Ff_L=\overline f_L$$
$$J_F \overline f_L=f_L$$
$$J_F f_R=\overline f_R$$
$$J_F \overline f_R=f_R.$$
Poincar\'e duality cannot be satisfied with this data as the intersection form is a $3\times 3$ antisymmetric matrix, and hence not invertible. However this does not affect the calculations above. 

The product geometries are candidates for the vacuum of the standard model. The axioms $D_F=D_F^*$,
$D_F\gamma_F+\gamma_F D_F=0$ and $D_F J_F=J_F D_F$ constrain the Dirac operator to the form
$$D_F=\begin{pmatrix}
0&M&G&0\\
M^*&0&0&H\\
G^*&0&0&\overline M\\
0&H^*&M^T&0\\
\end{pmatrix}$$
using the basis $(f_L,f_R,\overline f_L,\overline f_R)$, i.e. all the left-handed fermion basis vectors, followed by the right-handed ones, etc., where $G$ and $H$ are complex symmetric matrices. The first-order axiom constrains $M$, $G$ and $H$ further. According to the argument of Krajewski \cite{K}, $D_F$ can be decomposed by $D_F=D_L+D_R$, where $D_L$ commutes with the defining action (`left') action of $\cal A$ and $D_R$ commutes with the `right' action $Ja^*J^{-1}$. The structure of $D_F$ can thus be read off from the table of charges of the fermions. Let $(\lambda,q,m)\in \C\oplus\H\oplus M_3(\C)$. Then on each irreducible piece of ${\cal H}_F$ only one of these summands acts, as shown here. The particles in the table are the left-handed leptons, the right-handed electron, the left-handed quarks, the right-handed down quark and the right-handed up quark (and their generation partners).
$$\begin{matrix}
\text{particle}& \text{left action}&\text{right action}&\gamma_F \\
l_L&q&\lambda&1\\
e_R&\overline\lambda&\lambda&-1\\
q_l&q&m^T&1\\
d_R&\overline\lambda&m^T&-1\\
u_R&\lambda&m^T&-1\\
\overline l_L &\lambda&q^T&-1\\
\overline e_R&\lambda&\overline\lambda&1\\
\overline q_l&m&q^T&-1\\
\overline d_R&m&\overline\lambda&1\\
\overline u_R&m&\lambda&1\\
\end{matrix}$$
This means that the matrix $M$ has the same structure as in the Euclidean model: non-zero matrix elements connect the right-handed electron with the left-handed leptons, and the right-handed quarks with left-handed quarks. However the first-order condition forces $G$ to be zero. The matrix $H$ is allowed non-zero entries which connect $\overline u_R$ with $e_R$ (and similarly $u_R$ with $\overline e_R$). However these matrix entries are expected to be zero in the physical vacuum. This situation is somewhat similar to the possibility of lepto-quark terms in the Euclidean model \cite{leptoquarks}.

The model can be extended to include the right-handed neutrino, and the results are then rather different from the Euclidean case \cite{BD}. 
Following \cite{Schelp}, the table of charges is extended as follows.
$$\begin{matrix}
\text{particle}& \text{left action}&\text{right action}&\gamma_F \\
 \nu_R&\lambda&\lambda&-1\\
\overline\nu_R&\lambda&\lambda&1\\
\end{matrix}$$
The matrix $H$ is also extended and the first-order condition allows terms which connect $\nu_R$ with $\overline \nu_R$ (as well as with $\overline e_R$ and $\overline u_R$). 

\section{Fermionic action}

These terms in the Dirac operator can be interpreted in terms of the action of the standard model. For the Euclidean case, this action is the Connes-Chamseddine spectral action\cite{SAP}. Whilst it is not yet clear what the analogue of the bosonic part of the spectral action is for the Lorentzian case, the fermionic part 
$$ (\Psi,D \Psi)$$
can be used unchanged. To understand this formula, it is necessary to take account of the fact that fermionic fields are anticommuting Grassman variables in quantum field theory. To show that this formula gives the usual action, consider for example an electron field
$$\Psi=\psi_R\otimes e_R +  \overline\psi_R\otimes \overline e_R+
\psi_L\otimes e_L +  \overline\psi_L\otimes \overline e_L \in{\cal K}.$$
In this formula, the space-time field $\psi_R$ has right-handed chirality and $\psi_L$ left-handed, and the notation $\overline\psi_R$ is used for the charge conjugate $J_M\psi_R$.
The vectors $e_L$, $e_R$ etc., are taken to be unit vectors. 

Consider the action when the bosonic fields are the Minkowski space vacuum.
The action can be expanded, using the form of the Dirac operator for a product geometry,
$D=D_M\otimes 1+\gamma_M\otimes D_F$. The matrix $M$ is a scalar here, and so one can write 
 \begin{multline}\label{formula}(\Psi,D \Psi)= (\psi_R,D_M\psi_R)+(\overline\psi_R,D_M\overline\psi_R)+
(\psi_L,D_M\psi_L)+(\overline\psi_L,D_M\overline\psi_L)\\
-iM(\psi_L,\psi_R)+i\overline M (\overline\psi_L,\overline\psi_R)
+ i\overline M(\psi_R,\psi_L)-i M (\overline\psi_R,\overline\psi_L).\end{multline}
The space-time inner product $(\psi,\psi')=\int\overline\psi\psi'\;\dd^4x$ in this formula obeys the relation
$$(\psi,\psi')=(\overline \psi',\overline \psi)$$
once the anti-commutation of fermion fields is taken into account. This means that the fifth and eighth terms in (1) are equal, as are the sixth and seventh. 

In a similar way, the properties of the Dirac operator allow one to show that the kinetic terms are equal in pairs. The action reduces to 
$$\frac12 (\Psi,D \Psi)=(\psi_R,D_M\psi_R)+ (\psi_L,D_M\psi_L)
-i M(\psi_L,\psi_R)+i\overline M (\psi_R,\psi_L),$$
which is the usual formula for the Dirac action when $m=-iM$ is a real number.

For a neutrino field containing left- and right-handed neutrinos, 
$$\Psi=\psi_R\otimes \nu_R +  \overline\psi_R\otimes \overline \nu_R+
\psi_L\otimes \nu_L +  \overline\psi_L\otimes \overline \nu_L \in{\cal K},$$
the action has a Dirac kinetic and mass terms in the same way, but also an additional term
$$-i\overline H(\overline\psi_R,\psi_R)+ iH(\psi_R,\overline\psi_R).$$
This term is a Majorana mass term. If the three generations are taken into account, the action contains the terms
$$-i\overline H_{jk}(\overline\psi_R^j,\psi_R^k)+ iH_{jk}(\psi_R^j,\overline\psi_R^k),$$
summing over $j,k=1,2,3$.
Due to the fact that $(\psi_R^j,\overline\psi_R^k)=(\psi_R^k,\overline\psi_R^j)$, this formula
is compatible with the correct $3\times3$ symmetric matrix $H$ for generation mixing.
Therefore the Lorentzian model contains all the necessary ingredients for the see-saw mass generation mechanism for neutrinos. There is no Majorana term for the left-handed neutrinos.

\vskip .5in

\noindent\emph{Note added:} After finishing this work I learned that A. Connes has independently - and via different considerations -  arrived at the same idea of modifying the internal geometry to have signature (KO-dimension) 6 to solve the fermion doubling and neutrino mass problems \cite{CNEW}.


\begin{thebibliography}{999}


\bibitem[CC]{SAP} A.H. Chamseddine, A. Connes, \emph{The spectral
action principle.} Comm. Math. Phys. Vol.186 (1997),  731-750.

\bibitem[C]{C}
A. Connes, \emph{Gravity coupled with matter and the foundation of non-commutative geometry.} Comm.
Math. Phys.  182 (1996),   155--176.

\bibitem[CR]{CR} A. Connes,
\emph{Non-commutative geometry and reality.} J. Math. Phys. 36 (1995), 6194--6231.

\bibitem[S]{Schelp} R. Schelp, \emph{Fermion masses in non-commutative geometry.} Int. J. Mod.
Phys. B14, (2000), 2477-2484.

\bibitem[K]{K} T. Krajewski, \emph{Classification of finite
spectral triples.} J. Geom. Phys. Vol.28 (1998), 1-30.

\bibitem[BD]{BD} J.W. Barrett, R.A. Dawe Martins,
 Non-commutative geometry and the standard model vacuum.
  J.Math.Phys.47:052305 (2006) 

\bibitem[PSS]{leptoquarks} M. Paschke, F. Scheck, A. Sitarz, \emph{Can (noncommutative) geometry accommodate leptoquarks?} 
Phys.Rev.D59:035003 (1999)

\bibitem[BK]{BK}  Noncommutative geometry and the standard model of elementary particle physics.
Papers from the conference held in Hesselberg, March 14--19, 1999. Edited by F. Scheck, H. Upmeier and W. Werner. Lecture Notes in Physics, 596.
Springer-Verlag, Berlin, (2002)

\bibitem[LMMS]{LMMS} F. Lizzi, G. Mangano, G. Miele, G. Sparano. Fermion Hilbert space and fermion doubling in the noncommutative geometry approach to gauge theories.
  Phys.Rev.D55:6357-6366 (1997)

\bibitem[GIS]{GIS} J.M. Gracia-Bondia, B. Iochum, T. Sch\"ucker. The standard model in noncommutative geometry and fermion doubling.
   Phys.Lett.B416:123-128 (1998)

\bibitem[CNEW]{CNEW} A. Connes. Noncommutative geometry and the standard model with neutrino mixing, hep-th/0608226 (2006)
\end{thebibliography}
\end{document}